\documentclass[doublecol]{epl2}
\usepackage{graphics}
\title{Dimensional and temperature dependence of metal insulator transition
in correlated and disordered systems} 
\shorttitle{Disorder induced metal in 2D correlated system} 
\author{Tribikram Gupta\inst{1} \and Sanjay Gupta\inst{2} }
\shortauthor{Gupta \etal}

\institute{
  \inst{1} Institute of Mathematical Sciences, Chennai, 600113 - tgupta@imsc.res.in\\
  \inst{2} S. N. Bose National Centre for Basic Sciences, Kolkata, sanjay1@bose.res.in 
}
\pacs{71.10.Fd}{Lattice Fermion models}
\pacs{71.30.+h}{Metal-insulator transitions and other electronic transitions}
\pacs{71.23.An}{Theories and models;localized states}

\abstract{
We study the dimensional dependence of the interplay between
correlation and disorder in two dimension at half filling using  2D $t-t'$ 
disordered Hubbard model with deterministic disorder both at zero and finite temperatures.  
Inclusion of $t'$ without disorder leads to a metallic phase at half 
filling below a certain 
critical  value of $U$.  Above this critical value $U_c$ correlation favours 
antiferromagnetic phase. Since disorder leads to double 
occupancy over the lower energy site, the competition between Hubbard $U$ and disorder leads to the 
emergence of a metallic phase, which can be quantified by the 
calculation of Kubo conductivity, gap at half-filling , density of states, 
spin order parameter, Inverse participation ratio (IPR) and bandwidth.
We have  studied the effect of disorder on the system in a very 
novel way through a deterministic disorder which follows a Fibonacci sequence. 
Behaviour of different parameters show interesting features on going from a
two to quasi one dimensional system.}

\begin{document}






\maketitle

\section{Introduction}
Kravchenko  $et. al.$ \cite{Krav} in a series of experiments on very weakly 
disordered 2D semiconductors (Si MOSFETS) at very low filling, in the 
absence of magnetic field showed that it is possible for a 2D system to undergo Metal 
Insulator transition. Recently Punnoose and Finkelstein 
have shown that it is possible to identify a Quantum Critical point  
separating a metallic phase stabilized by electronic 
interactions from the insulating phase where disorder prevails over 
electronic interactions in certain systems \cite{Punnoose}. 

Recently the Mott Insulator has been realised
for fermionic atoms in an optical lattice in 3 dimensions\cite{Jordens}
whereas most of the previous results were for Bosonic atoms.
Further the experimental realisation of Anderson insulators for Bose
gas has been realised\cite{Damski} by disordering the optical lattice 
site potentials. It is possible to disorder the lattice in a 
rather controlled manner on optical lattices. This has facilitated 
the controlled experimental study of the phase diagram arising due to the 
competition of strong correlations and disorder in these systems, causing us   
to revisit this topic. 
 
The present work emphasizes on the dimensional dependence of the interplay between
electronic correlation and deterministic disorder 
(and the consequential reflection in metal insulator transition(MIT)) 
as one goes from a two dimensional system to a quasi one dimensional system. 
Also the important aspect of the role of temperature in this whole scenario 
has been duly addressed.

A Hartree Fock mean field treatment of the non-disordered $t-t'$ Hubbard 
Hamiltonian \cite{Hirsch} shows Paramagnetic 
to Antiferromagnetic transition at $U_c$ = 2.1.e The disordered  
Hubbard hamiltonian was shown to have persistent currents \cite{Bouzerar}, 
Milovanovi\'{c} et al \cite{milovanovicetal89} investigated it numerically 
and Bhatt et al \cite{bhattfisher92} studied local moment formation. 
Dobrosavljevi\'{c} and co-workers 
\cite{EffMediumDMFT} use generalized DMFT method 
to report the interesting fact that increasing disorder from a clean Mott
insulator results in the closing of the Mott gap and the stabilization
of a metallic phase.

The possibility that a metallic phase may exist in 2D due to the interplay of 
strong correlations and random disorder for the half filled disordered Hubbard Model has 
been suggested \cite{Nandini}.

We solve the problem numerically using Unrestricted Hartree Fock(UHF) 
technique. The UHF method works remarkably well even in 1D at half-filling and is
 in excellent agreement with results obtained from real space renormalization method
 \cite{Gupta}. In 2D at half filling the UHF method gives result in close proximity with
that of quantum monte carlo method \cite{Hirsch}.
However, one should mention here that the correlated Kondo like processes that 
lead to large effective-mass
renormalization in the vicinity of the Mott transition can only be captured 
by a method like DMFT and not by the present method \cite{EffMassRenorm}. 
The importance of inelastic scattering processes for calculating the conductivity
was shown by a DMFT calculation by Aguiar {\it et. al} \cite{Aguiar}.
 We choose a deterministic binary disorder(that follows a Fibonacci sequence),  
motivated largely by optical lattice realisations. 
Deterministic disorder has been one of the forefront areas of research 
in condensed matter physics for the past two  and a half decades both in experiments
 \cite{exptDetDis}
and theory \cite{TheoryDetDis}.
Theoretical studies of the quasi-periodic systems have shown that in
 the non-interacting 
limit the wave function shows a power law localization both in one \cite{1Dpowerlaw} 
and two \cite{2Dpowerlaw} dimensions. Very recent experiment in 1D quasi-periodic 
optical lattice \cite{NatureOptLat} where the system was described by Aubry-Andre Hamiltonian
has shown exponentially localized states (Anderson localization) in the large disorder limit. A theoretical explanation
has been published recently \cite{NJP} using a Fibonacci sequence in 1D.  
 We benchmark our
results and trends against old results obtained with random disorder.
Zero and finite temperature calculations have been carried out for different 
values of the disorder strength. The impurity concentration remained almost 
constant as we reduced the dimension, which is an artifact of the Fibonacci 
sequence.  
We have taken a $40\times40$ lattice which is fairly large to rule 
out finite size  effects.

The system, though insulating in the $U > U_c$ regime, undergoes a depletion
of the charge gap at half filling as the disorder is increased, 
 till it reaches a stage where the lower energy sites would be 
doubly occupied in spite of the large value of $U$. A narrow metallic 
regime emerges while going from the Neel state to the state with a distribution
of doubly-occupied and unoccupied sites.
The low temperature metallic 
phase gives way to an insulating phase upon heating. 
It has residual spin order(very low value) and is thus not a perfect paramagnet. 
One of our objectives in the present work is to investigate the evolution 
of the interplay between disorder and interactions and relatively whose effect gets 
more dominant as one reduces the dimension. 
The effect of next nearest neighbour hopping makes the delocalized phase more robust.  

\section{Disordered Hubbard Hamiltonian}
The disordered 2D t-t' Hubbard Hamiltonian considered by us is as follows:
\begin{eqnarray} 
\label{eq.1}
\it{H_{o}}=\sum_{i}\epsilon_{i}n_{i\sigma}
-t\sum_{i}(c^{+}_{i\sigma}c_{i+{\delta_1}\sigma}+h.c) \nonumber \\ 
+t'\sum_{i}(c^{+}_{i\sigma}c_{i+\delta_2\sigma}+h.c) 
\end{eqnarray}
\begin{equation} 
\label{eq.2}
\it {H_{int}}=U\sum_{i}n_{i\uparrow}n_{i\downarrow}
\end{equation}
\begin{equation} 
\label{eq.3}
\it {H}=\it{H_{o}}+\it{H_{inc}}
\end{equation}
\noindent
Here $t$ and $t'$ are nearest and next nearest 
neighbour hopping terms, $U$ is the on site Hubbard interaction, $\epsilon_{i}$ is 
the site potential at i-th site, while $c^{+}_{i\sigma}$~, $c_{i\sigma}$~, 
$n_{i\uparrow}$~, $n_{i\downarrow}$ 
are creation, annihilation operators for the electron of spin $\sigma$,
and the number operators for up and down spins respectively at site $i$. 

\section{Deterministic Disorder}
Deterministic disorder is neither periodic nor fully random and  
have long range correlations(correlated disorder). 
The site potential  $\epsilon_{i}$ is deterministically disordered. It follows a 
Fibonacci sequence which is generated as
$A \rightarrow AB$ and $B \rightarrow A$, where $A$ and $B$ are the two different sites.
In our case site $B$ has a higher site potential than site $A$ ($\epsilon_B > \epsilon_A$,
$W=\epsilon_B-\epsilon_A$, being the disorder strength).
 A typical Fibonacci chain in a particular generation looks
like: $ABAABABAABAABABAABABA......$. 
We have used the Fibonacci sequence and generalised
the idea of quasi-periodicity in one dimensions to two dimension by 
essentially wrapping up the one dimensional Fibonacci sequence over 
the two dimensional square lattice. Moreover the
number of lattice sites need not be a Fibonacci number.
In our case the $N$ lattice sites picks up the first
$N$ entries of the sequence from the left, thus allocating the site potentials. 
This sort of disorder can also be referred to as correlated disorder. 
 
\begin{table}
\caption{Number of A $\&$ B sites for different system sizes.}
\label{tab.1}
\begin{center}
\begin{tabular}{l c r r}
size          & A   &  B  &  $A/B$\\
$40\times40$  & 989 & 611 & 1.608\\
$40\times14$  & 346 & 214 & 1.616\\
$40\times10$  & 247 & 153 & 1.614\\
$40\times6$   & 148 & 92 & 1.608
\end{tabular}
\end{center}
\end{table}

\section{Discussion of Numerical Method}
We solve the Hartree Fock decoupled Hamiltonian with an initial guess(seed) 
of $n_{i\uparrow}$ and $n_{i\downarrow}$, and solve it self-consistently 
till the solutions of $n_{i\uparrow}$ and $n_{i\downarrow}$ converge
to a difference of less than $10 ^ {-7}$ for every site.
We settled up with that initial seed which minimises the ground state energy.
This strict convergence 
criterion has been employed to ensure that we stay 
away from local minimas while exploring the energy 
landscape. This is extremely important as we approach the disorder induced 
localized phase using a variational approach like Hartree Fock to treat interactions
as our only guiding principle in choosing the ground state configuration is 
energetics. In the low and medium disorder regime if we start from the 
Neel order state as the starting seed, we get the desired convergence very quickly, 
while in the high disorder regime near the MIT, we have 
found the desired results starting with a near paramagnetic configuration as the 
initial seed(PM seed).  

\subsection{Unrestricted Hartree Fock }
In UHF theory the Hubbard interaction term is decoupled keeping in 
the site dependence and thus one obtains modified site potentials for up 
and down spins which reads as: 

\begin{equation}
\label{eq.4}
\epsilon^{'}_{i\downarrow}=\epsilon_{i}+U<n_{i\uparrow}>; 
\epsilon^{'}_{i\uparrow}=\epsilon_{i}+U<n_{i\downarrow}> 
\end{equation}
This gives us the following Hamiltonian which is decoupled 
into up and down spin parts: 
\begin{eqnarray}
\label{eq.5}
\it {H_{\sigma}}=\sum_{i}\epsilon^{'}_{i\sigma}n_{i\sigma}-
t\sum_{i}(c^{+}_{i\sigma}c_{i+\delta_1\sigma}+h.c) \nonumber \\ 
+t'\sum_{i}(c^{+}_{i\sigma}c_{i+\delta_2\sigma}+h.c)
\end{eqnarray}
Where $\epsilon^{'}_{i\sigma}$ are now defined by eq.~(\ref{eq.4}).

\section{Quantities calculated}
Once the energy spectrum is obtained the 
energy gap at half-filling is calculated by the difference between the energy 
level at half-filling and the one just above it. The expression reads as: 
\begin{equation}
\label{eq.6}
E_{gap}=E_{N/2+1} - E_{N/2}
\end{equation}
where $N$ is the number of sites in the 2D lattice. The ground state 
energy $E_g$ is calculated by summing up the single particle energy states 
for both spin species up to the Fermi level corresponding to the desired 
filling. The Fermi level at half filling is calculated as the mean of the 
highest filled state and the lowest unoccupied one to ensure that the system 
remains half filled. 
All finite temperature calculations are performed with respect to this Fermi 
energy with the appropriate Fermi distribution factors. 
If $n_{i\uparrow}$ and $n_{i\downarrow}$ are converged values of
of the occupation numbers of up and down spins respectively at the
i-th site, then the spin order is defined as:
\begin{equation}
\label{eq.7}
S={1\over N}(\hskip -.1in\sum_{i,evenrow}\hskip -.05in(-1)^{i}(n_{i\uparrow}-n_{i\downarrow})
+\sum_{i,oddrow}\hskip -.05in(-1)^{(i+1)}(n_{i\uparrow}-n_{i\downarrow}))
\end{equation}
	For disordered non interacting systems, the Kubo formula, at any
	temperature is given by:
\begin{equation}
\label{eq.8}
	\sigma ( \omega)
	= {A \over N}
	\sum_{\alpha, \beta}(n_{\alpha}-n_{\beta})
	{ {\vert f_{\alpha \beta} \vert^2} \over {\epsilon_{\beta}
	- \epsilon_{\alpha}}}
	\delta(\omega - (\epsilon_{\beta} - \epsilon_{\alpha}))
\end{equation}
with $A = {\pi  e^2 }/{{\hbar a_0}}$, $a_0$ being the lattice spacing,
and $n_{\alpha}$ = Fermi function with energy $\epsilon_{\alpha}-\mu$.
The $f_{\alpha \beta}$ are matrix elements of the current operator
 $j_x = i t  \sum_{it, \sigma} (c^{\dagger}_{{i + x a_0},\sigma}
 c_{i, \sigma} - h.c)$, between exact single particle eigenstates
 $\vert \psi_{\alpha}\rangle$,
 $\vert \psi_{\beta}\rangle$, {\it  etc},  and
 $\epsilon_{\alpha}$, $\epsilon_{\beta}$
 are the corresponding eigenvalues. In this paper, conductivity/conductance
 is expressed in units of $A = {\pi  e^2 }/{{\hbar a_0}}$.

We calculate the `average' conductivity over 3-4 small frequency intervals 
$\Delta \omega$($\Delta \omega$ = $n \omega_{r}$, n = 1,2,3,4),
and then differentiate the integrated conductivity to get $\sigma(\omega)$
at $\omega = n \omega_{r}$, n = 1,2,3. These three values are then extrapolated 
to $\omega = 0$ to get $\sigma_{dc}$. For the sake of simplicity, and keeping 
all other consistency checks in mind \cite{SanjeevEPL}, we have taken the value of 
$\sigma(\omega_r)$ as the actual value of $\sigma_{dc}$.  

The average of $\sigma(\omega)$  over the
interval $[0, \omega_r]$ is 
\begin{equation}
\label{eq.9}
	\sigma_{int}(\omega_{r}, \mu,  N)
	= {1 \over {\omega_{r}}}\int_0^{\omega_{r}}
	\sigma(\omega, \mu, N)d \omega
\end{equation}
	$\omega_{r}$ is set to be sufficiently larger than the average spacing between 
	energy levels, given by $D/N$, where $D$ is the full bandwidth for the interacting 
	2D disordered Hubbard Model that we are considering and is simply calculated as
        $D = E_{N} - E_{1}$, where $E_{N}$ and $E_{1}$ refer to the topmost and lowermost eigenvalues. 
         For $N = 1600$, $\omega_{r} = 0.05$. 

	In the discussion below, all length scales are normalized by the lattice 
	parameter $a$, and we use dimensionless energy parameters $U$,  
	$W$ and $t'$, scaled by the hopping amplitude $t$.  

       IPR which gives the measure of how many sites are participating in a
        particular eigenfunction is defined as following \cite{Adame}:\\
\begin{equation}
\label{eq.10}
      IPR(\ E_{n})=\sum_{l=1}^{L}|\phi^{(n)}_{l}|^{4}/(\sum_{l=1}^{L}|\phi^{(n)}_{l}|^{2})^2
\end{equation}
      where $l$ corresponds to the site index,$E_{n}$ is the nth eigenvalue. 
      The cited ref calculated IPR for a non interacting system with
      Fibonacci modulated site potential.
      It is clearly visible from the above expression that larger value of
      IPR corresponds to localized electronic states whereas small value
      of IPR indicates an extended state. In our case the topmost eigenfunction
      near the Fermi level was considered for this purpose.           

\section{Numerical Results}

	We have calculated the gap at half filling, charge and spin order, bandwidth, 
	IPR and the $\sigma_{dc}^{xx}$  
	of the 2D $t-t'$ disordered Hubbard model at half filling as we vary the 
	$t'$, Disorder potential $W$ and also reduce the 
	dimensionality along the y direction. We study the problem for intermediate 
	disorder strength ($W < U$) where the underlying Neel magnetic ordering(due 
	to $U$) is not killed and is predominantly present, for a $40\times40$ lattice.
	We have shown here our calculations for system sizes $40\times40, 40\times14, 
	40\times10$ and $40\times6$. 
	We have tabulated the number of $A$ and $B$ sites and the ratio $A/B$
	for the different sizes which is given in table~\ref{tab.1}

\begin{figure}
  \begin{center}
    \begin{tabular}{cc}
      \resizebox{39.5mm}{!}{\includegraphics{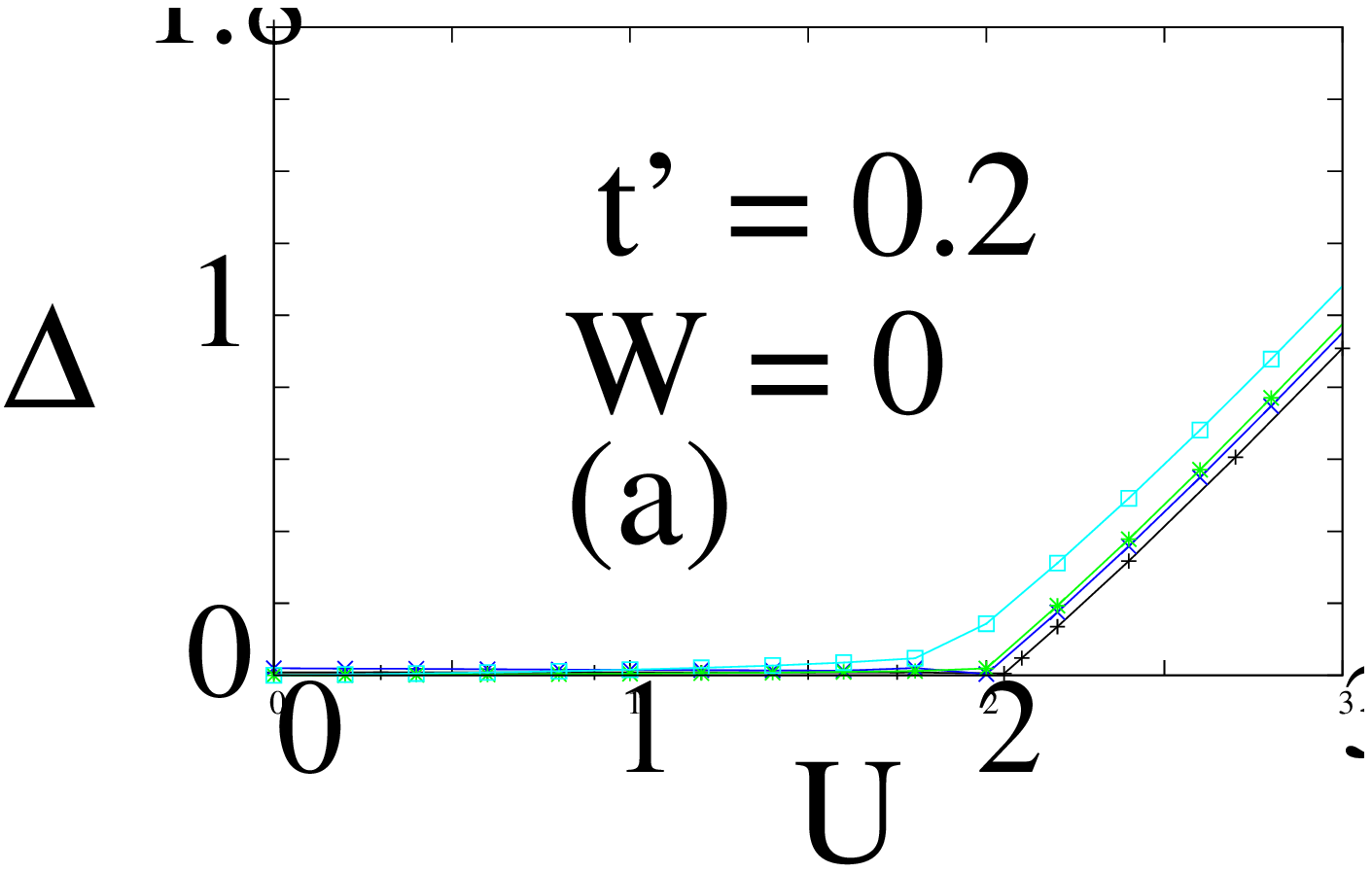}} & 
      \resizebox{39.5mm}{!}{\includegraphics{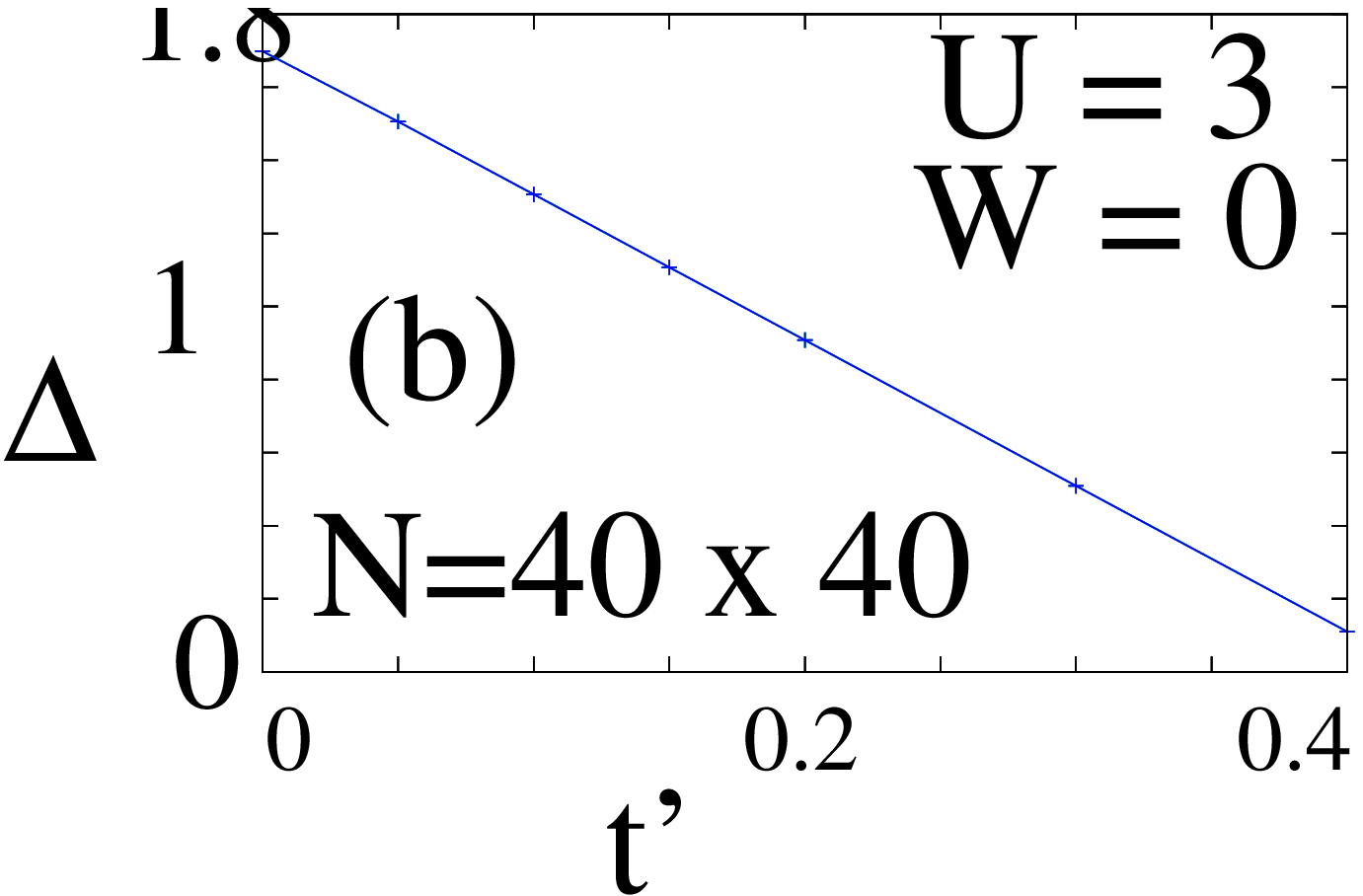}} \\
     \end{tabular}
\caption{Fig.~\ref{fig.1}a shows the effect of reducing $n_y$ on the gap at half
filling. For $t'=0.2$, gap is zero till $U_c=2.1$. $U_c$ is unchanged
all the way down to $40\times14$ below which $U_c$ starts decreasing slowly at first,
and below $40\times8$, it shows a steep fall, as the quasi 1D effect
sets in and electronic motion gets constrained. Fig.~\ref{fig.1}b shows the fall in gap
at half filling with increasing $t'$. The plot is for $W = 0$, $U = 3$
and $N = 40\times40$}
\label{fig.1}
\end{center}
\end{figure}

%
%

\begin{figure}
  \begin{center}
    \begin{tabular}{cc}
      \resizebox{39.5mm}{!}{\includegraphics{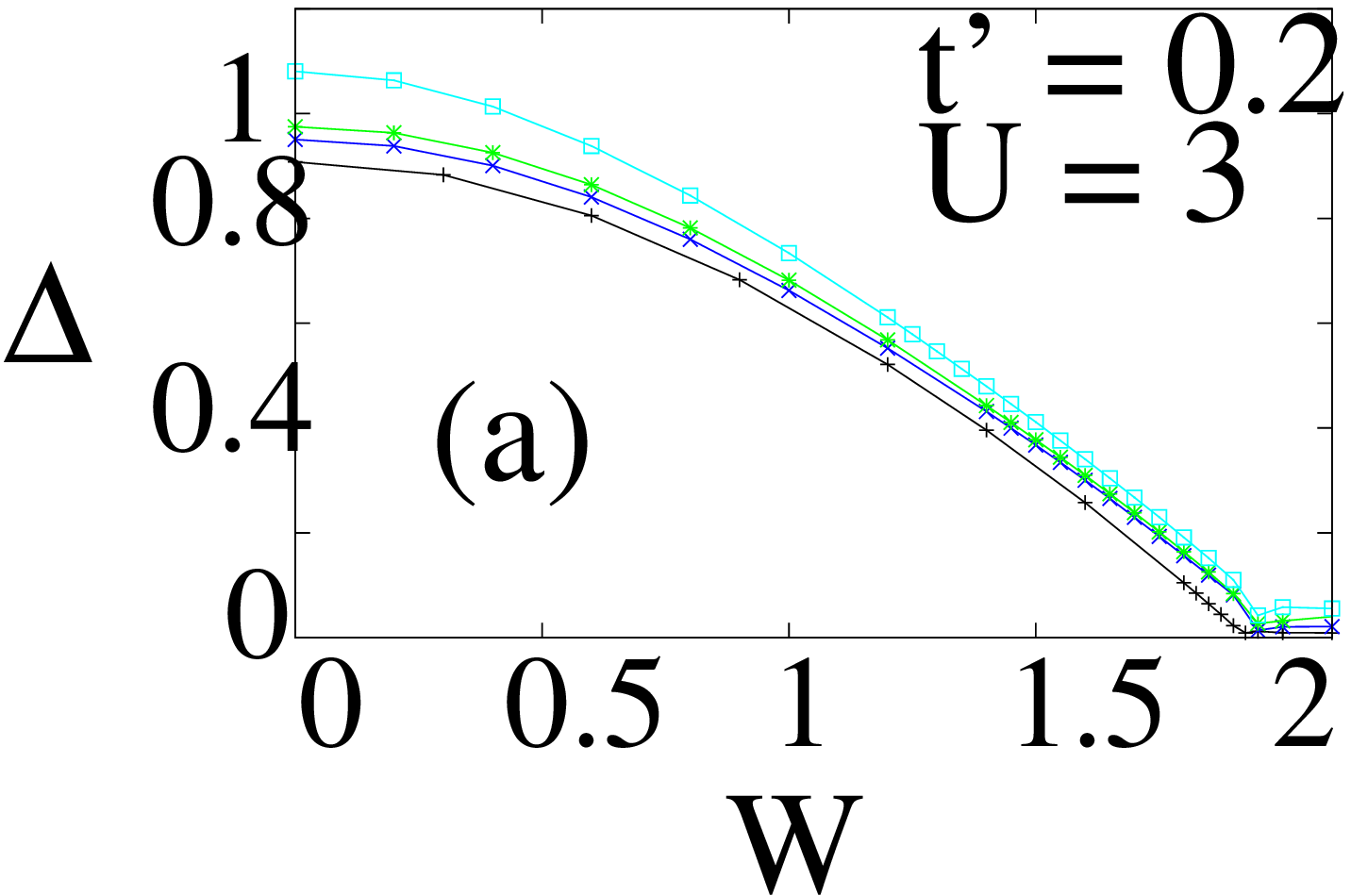}} & 
      \resizebox{39.5mm}{!}{\includegraphics{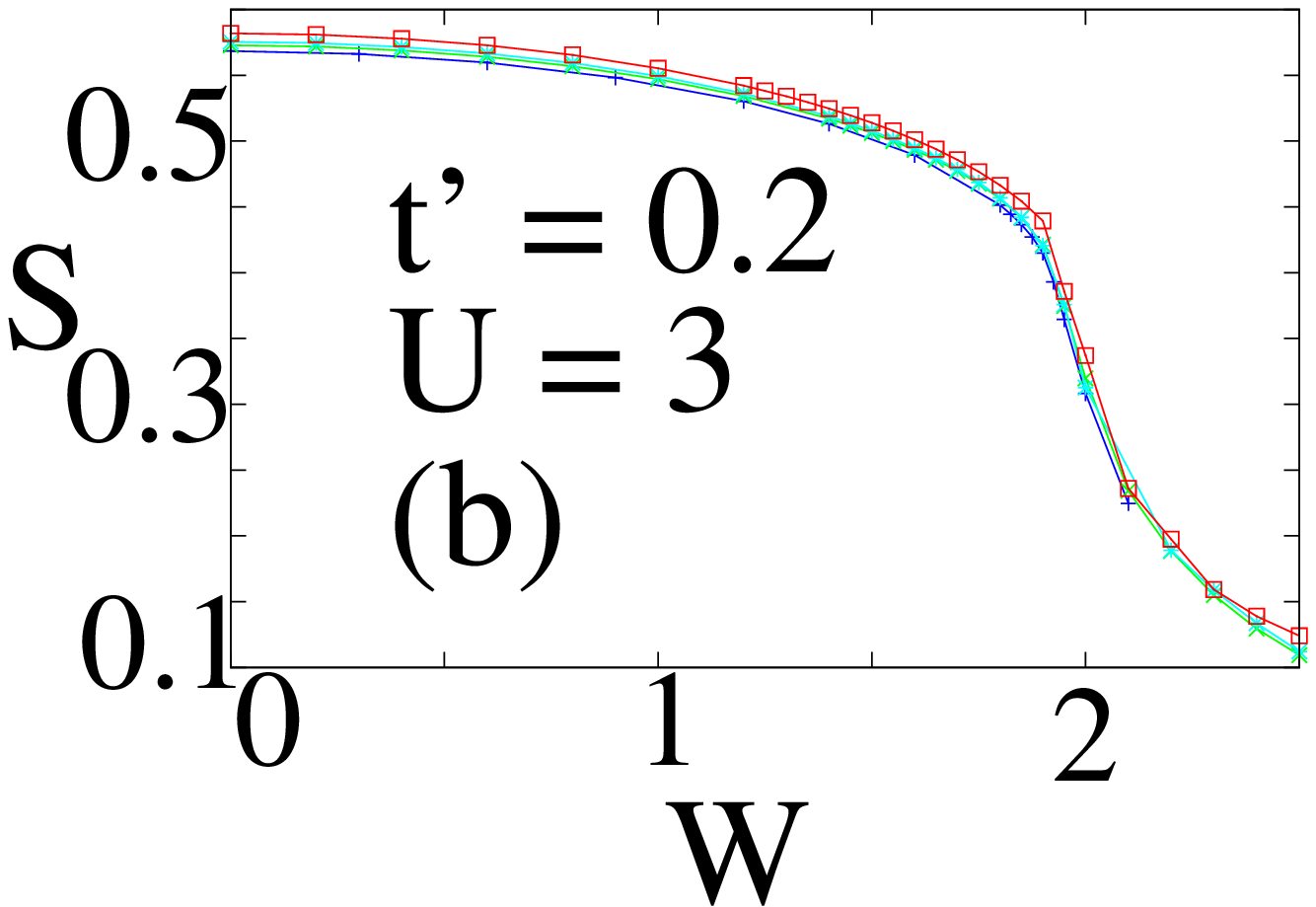}} \\
     \end{tabular}
\caption{Fig.~\ref{fig.2}a shows the plots of the gap at half filling and Fig.~\ref{fig.2}b shows 
the plots of spin order for $t'=0.2$, $U=3.0$ as a function of disorder strength $W$,
for various values of $n_y$.}
\label{fig.2}
\end{center}
\end{figure}

	In Fig.~\ref{fig.1}a we study effect of size reduction along y axis. 
	We show that for $t'=0.2$, $E_{gap}$ is zero till a critical 
	$U_c=2.1$, above which it becomes non-zero and matches well with 
	Hirsch's result\cite{Hirsch}. $U_c$ is unchanged all the way down 
	to $40\times14$ below which $U_c$ starts decreasing slowly at first, 
	and below $40\times8$, it shows a steep fall, as the quasi 1D effect 
	sets in and electronic motion gets constrained.  In Fig.~\ref{fig.1}b $E_g$ 
	decreases almost linearly as we increase $t'$, in the 
	regime( $U > U_c, U_c = 2.1$), as $t'$ kills the Neel ordering due
	to $U$.


	Fig.~\ref{fig.2}a shows the plot of the gap (as we reduce $n_y$) for $t'=0.2$ 
	against disorder strength $W$ 
	for $U=3.0$, which renders the undisordered phase antiferromagnetic. 
	The disorder reduces the gap as now the electrons of both spin species 
	will try to avoid the site with higher energy, thereby allowing some double 
	occupancies on the lower energy site. The gap decreases rather slowly for low 
	values of $W$ and finally falls sharply at a certain $W = W_c$, for a fixed system 
	size. 

	We find that for a fixed disorder, though $E_g$ 
	increases as we reduce $n_y$, the rate of fall in $E_g$ increases
	with increase in disorder as we reduce $n_y$, highlighting   
	that effect of disorder is more pronounced than the effect of Hubbard $U$ 
	as one reduces $n_y$. 
	We will see the effect of this in our Kubo conductivity results.

	The corresponding spin order plot also follows a similar trend (Fig.~\ref{fig.2}b). 
        However the 
	reduction in this case is more gradual than the gap at half filling. This is because
	the system prefers to retain the antiferromagnetic configuration to gain maximum 
	kinetic energy due to hopping, till a critical value of $W$ upto which $U$ 
	is effective, and beyond which the spin order reduces very rapidly.
	We find this critical $W \simeq 1.75$, at which point the Hubbard gap roughly 
	diminishes from first to second place of decimal. The reduction in magnetic
	ordering is much slower than the gap at half filling, and there is significant 
	residual magnetic ordering($S \simeq 0.25$) when $\Delta$ has already gone to zero.

\begin{figure}
  \begin{center}
    \begin{tabular}{cc}
      \resizebox{39.5mm}{!}{\includegraphics{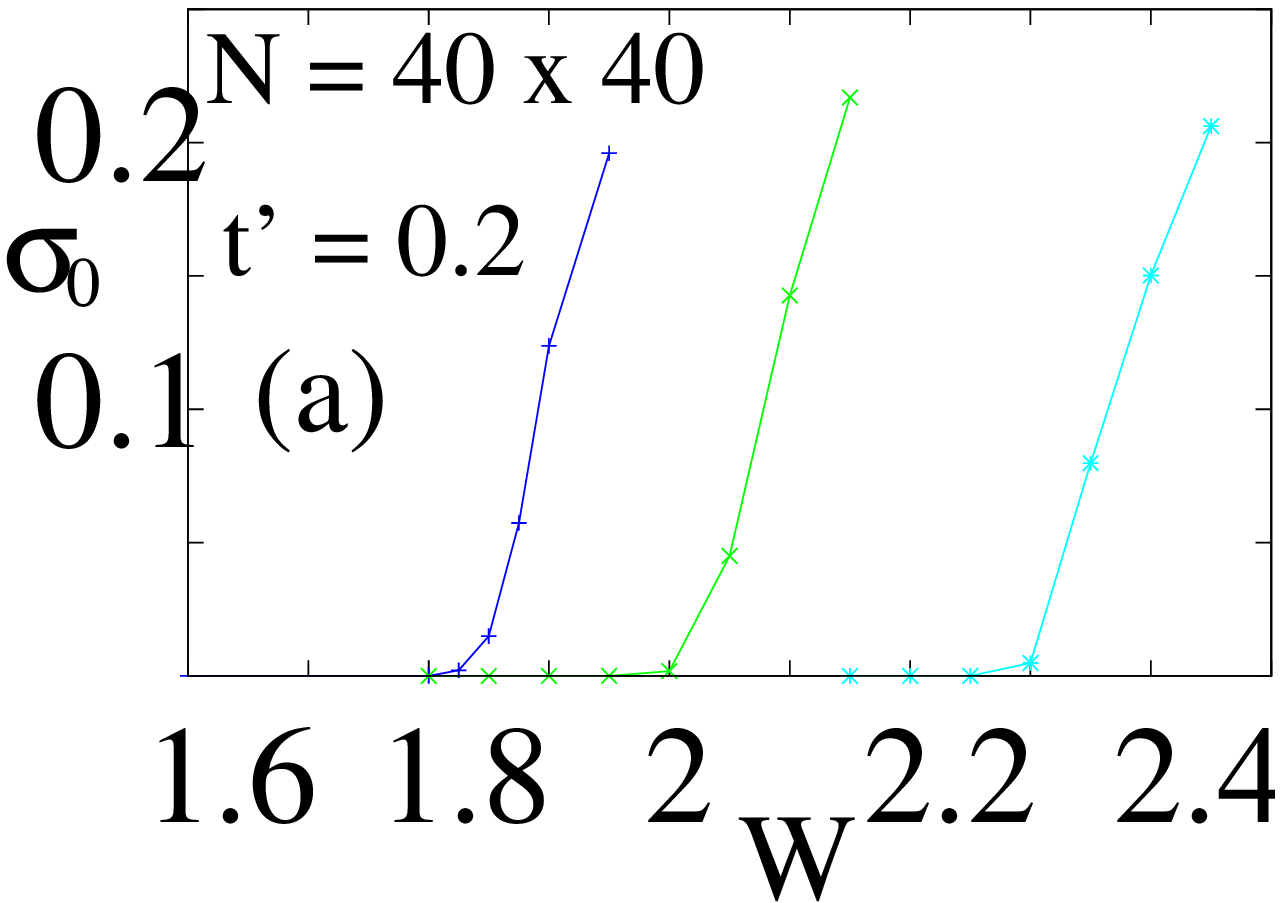}} & 
      \resizebox{39.5mm}{!}{\includegraphics{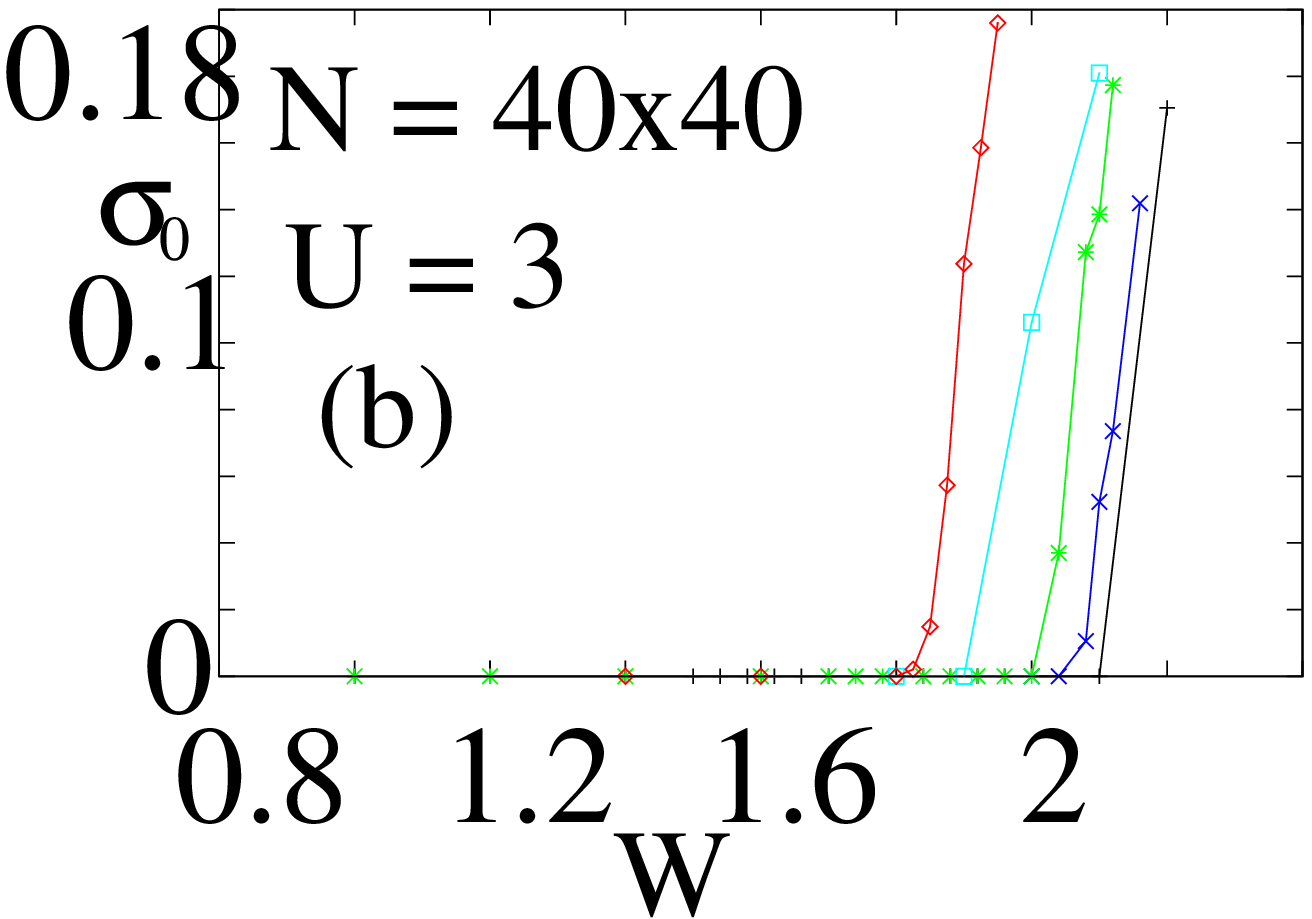}} \\
      \resizebox{39.5mm}{!}{\includegraphics{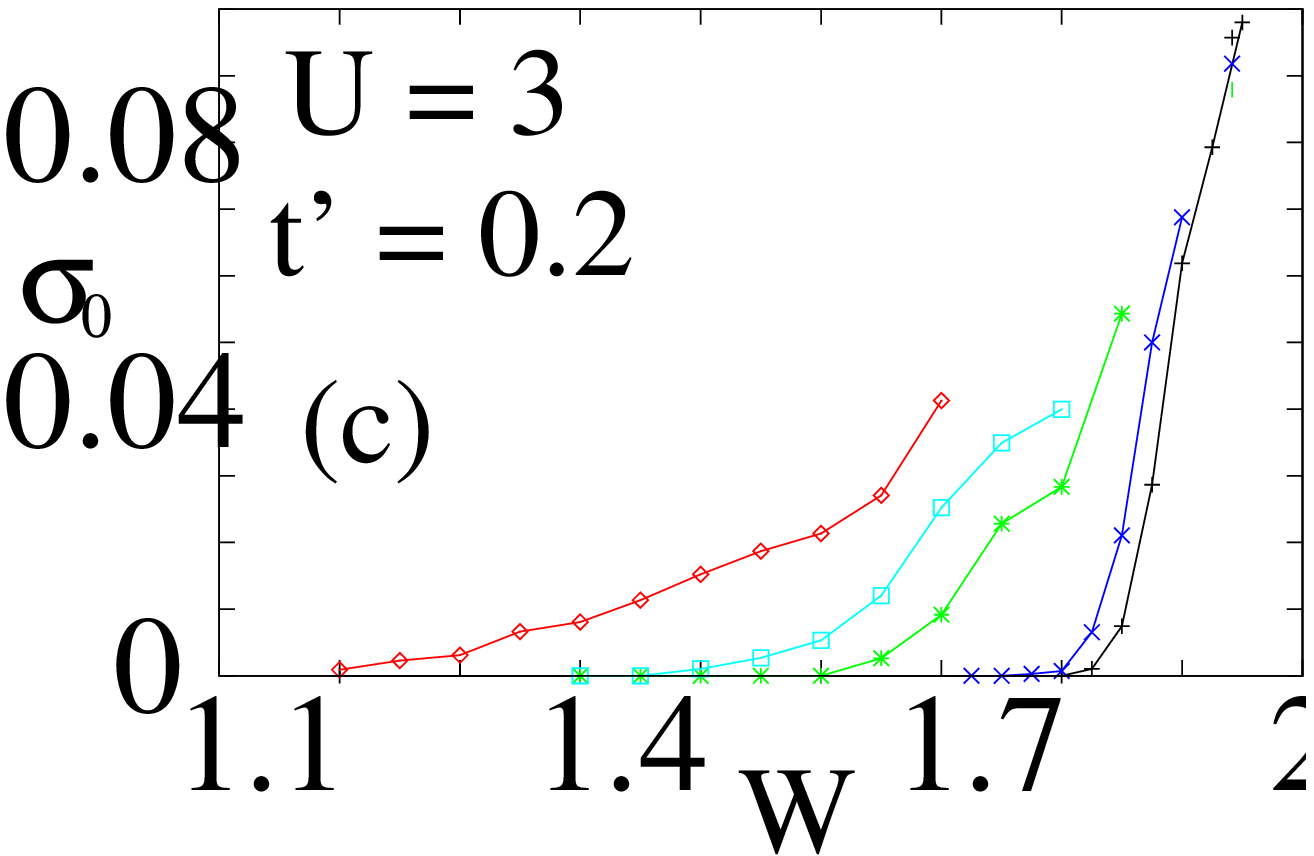}} &
      \resizebox{39.5mm}{!}{\includegraphics{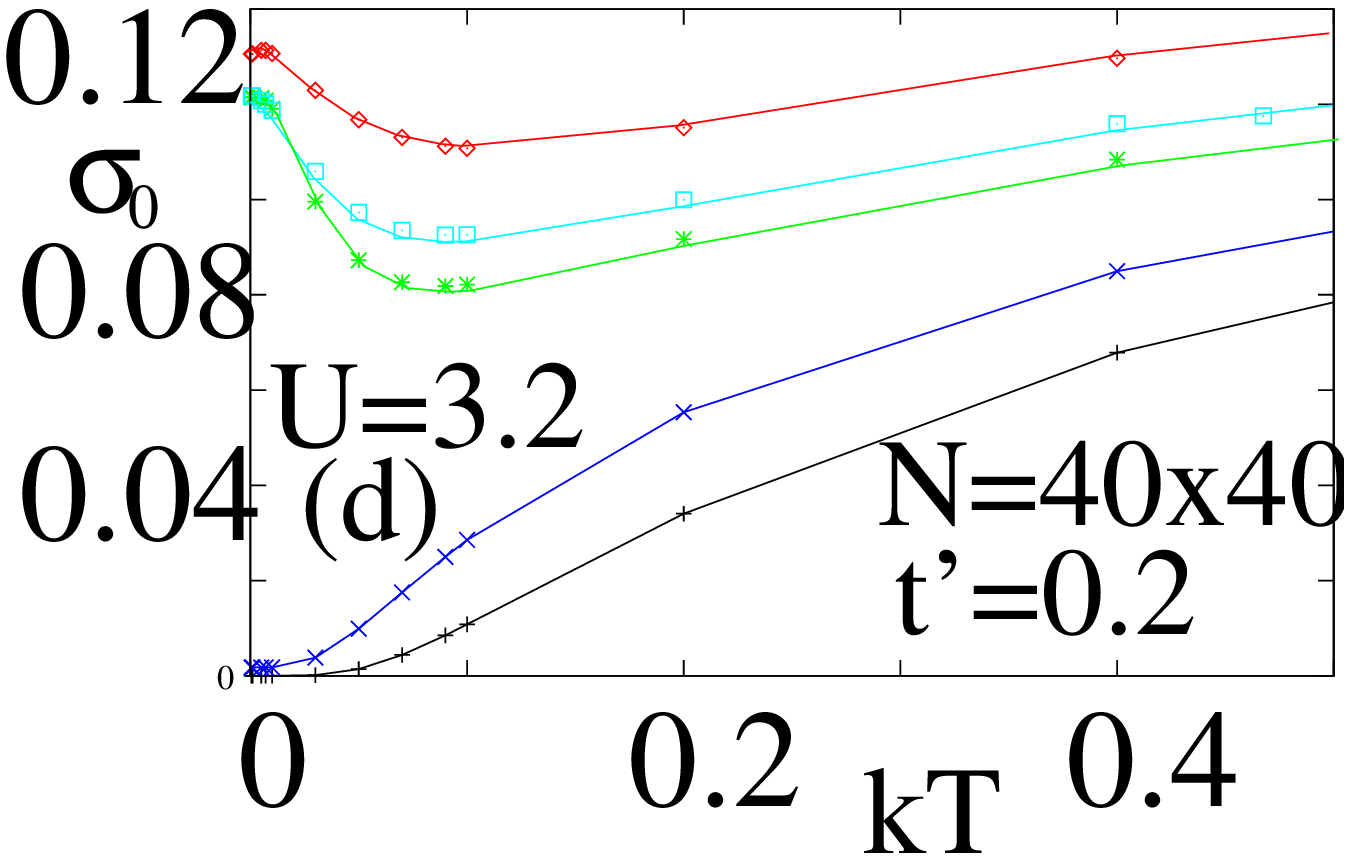}} \\
      \resizebox{39.5mm}{!}{\includegraphics{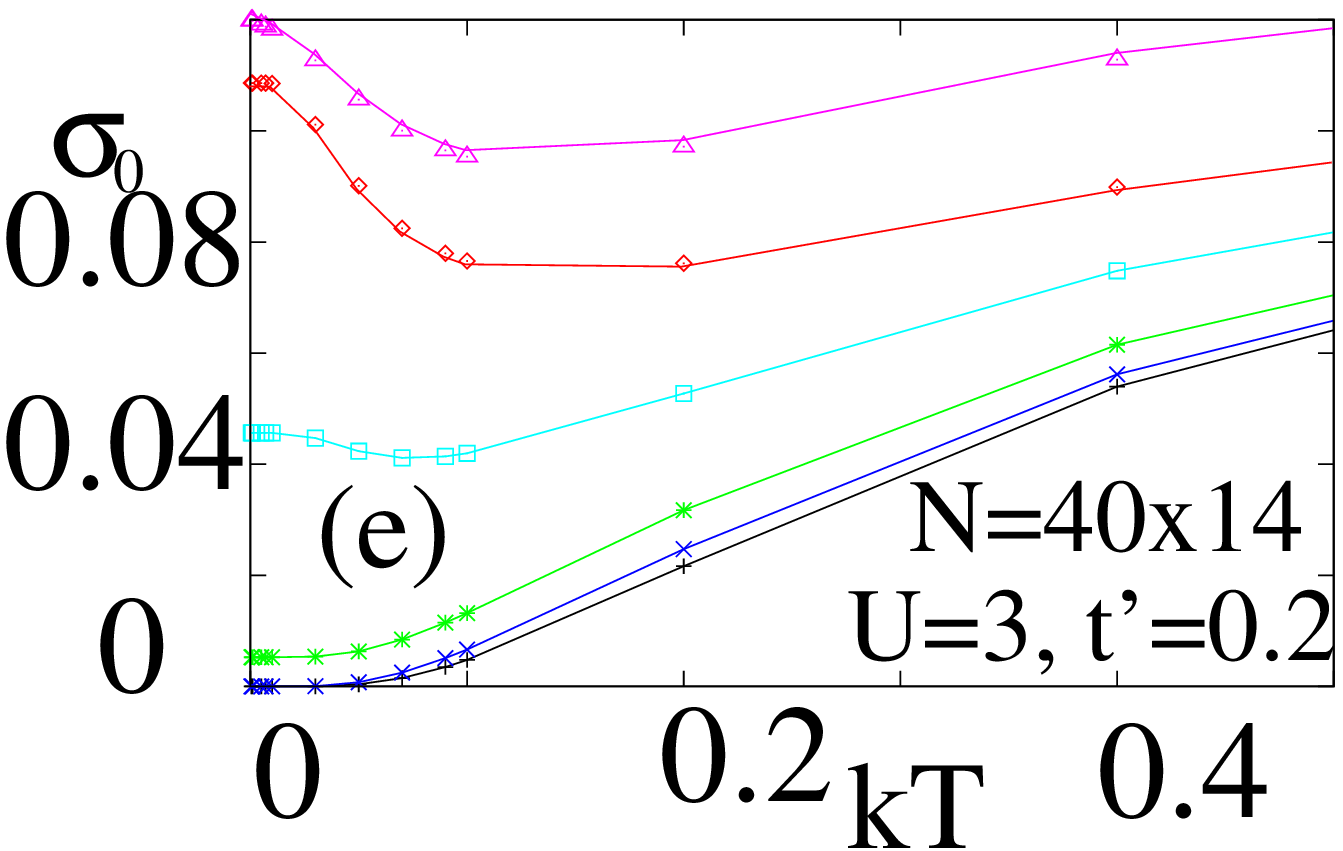}} &
      \resizebox{39.5mm}{!}{\includegraphics{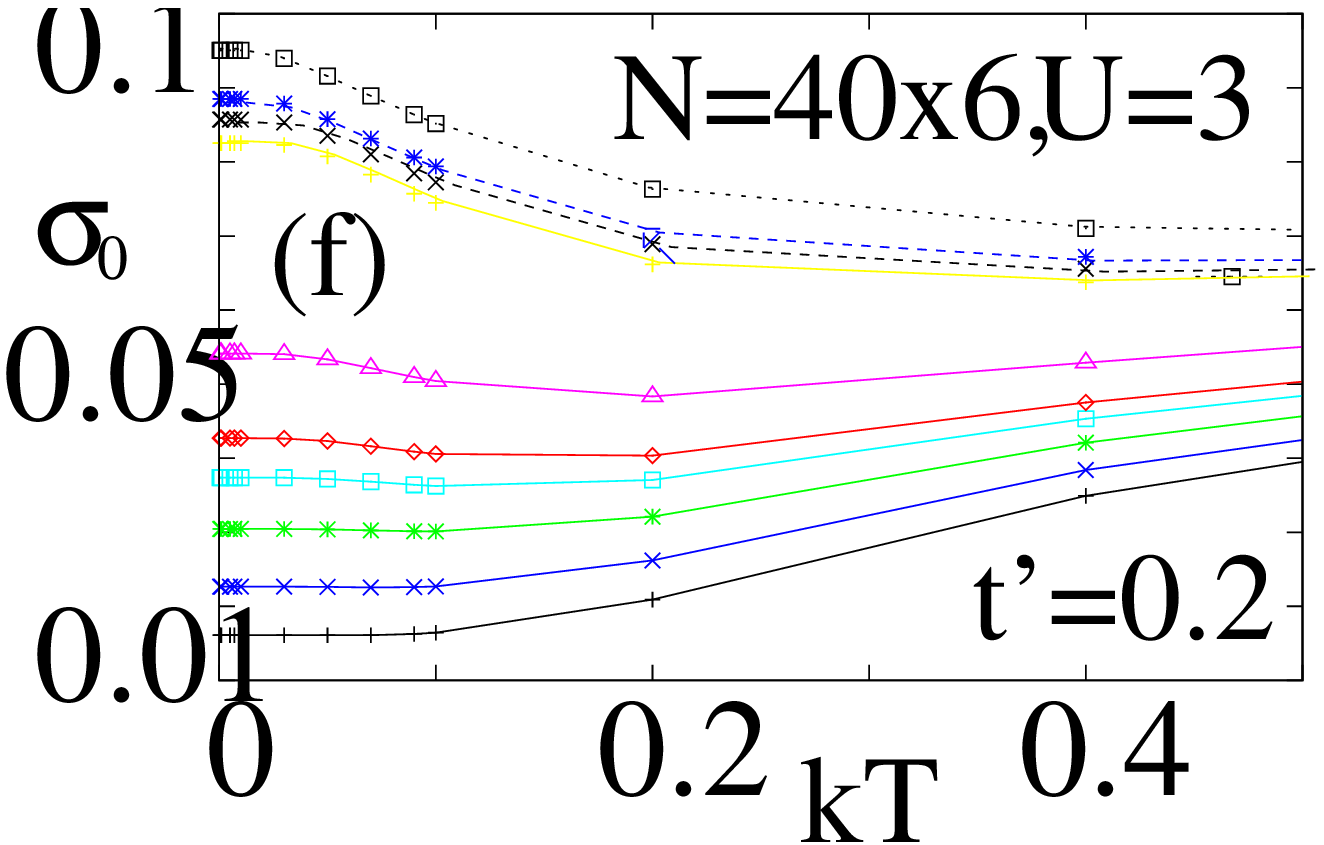}} \\
     \end{tabular}
\caption{Fig.~\ref{fig.3}a,b,c are plots of Kubo conductivity against disorder. 
Fig.~\ref{fig.3}d,e,f are plots of Kubo conductivity against temperature. 
In Fig.~\ref{fig.3}a plots are for $U$ = 3,3.2,3.4
from left to right. The Kubo conductivity becomes non-zero and system becomes metallic
for higher values of $W$ as $U$ increases.
The curves in Fig.~\ref{fig.3}b are for $t'$ = 0.2, 0.15, 0.1, 0.05, 0 from left to right.
Fig.~\ref{fig.3}c shows the effect of reducing $n_y$ on the Kubo conductivity from left to
right, $n_y$ = 6,10,14,30,40. Fig.~\ref{fig.3}d shows MIT at low temp driven by increasing
disorder. The curves correspond to $W$ = 1.8,2,2.1,2.15,2.2 from bottom to top.
Fig.~\ref{fig.3}e and Fig.~\ref{fig.3}f are for $n_y$ = 14 and 6 respectively and highlight how the
MIT takes place at lower value of disorder as we reduce $n_y$. The curves in Fig.~\ref{fig.3}e
are for $W$ = 1.45 to 1.95 in steps of 0.1 from bottom to top, while Fig.~\ref{fig.3}f are for
$W$ = 1.4, 1.85 in steps of 0.05 from bottom to top.}
\label{fig.3}
\end{center}
\end{figure}

	Fig.~\ref{fig.3}a shows the plot of the dc conductivity($\sigma_0$)  
	against disorder strength $W$ at $T$ = 0 for $U$ = 3,3.2 
	and 3.4, which is well in the 
	antiferromagnetic regime even for $t'$ = 0.2. Due to interplay between 
	correlation and disorder, $\sigma_0$ increases from zero at a critical 
	value of $W = W_c$, whose value increases with $U$. Thus  
	it takes higher values of disorder to form doubly occupied 
	sites as one increases $U$ at half filling. This plot cannot be 
	extrapolated indefinitely on the lower side as $U$ has to be 
	greater than $U_c$. 

	Fig.~\ref{fig.3}b shows the plot of Kubo conductivity at zero temperature as we change 
	$t'$ for the disordered problem. It is found that as we increase $t'$ from zero 
	it takes a lower value of disorder to first generate a non zero value of the 
	Kubo conductivity and then render the system metallic. 

	Fig.~\ref{fig.3}c shows the plot of the $T = 0$ $\sigma_0$ against $W$ as we reduce 
	the system size along y direction, thus going over to a quasi-1D system.  
	We need lower disorder value to annul the effect of $U$, as we lower the 
	dimensionality again highlighting the fact that the effect of disorder, 
	for a fixed value of $U$ becomes more robust as we decrease the 
	dimensionality from 2D to quasi 1D. 
	The faster rate of depletion of the Hubbard gap $\Delta$ as we reduce the
	dimensionality as seen in Fig.~\ref{fig.2}a, is thus connected to the $\sigma_{0}-W$
	curves for different $n_y$ in this way. 

	Fig.~\ref{fig.3}d show the plots of  $\sigma_0$ against  
	temperature for $U =  3.2$ and shows an insulator to metal transition
	with increasing $W$ at low temperatures. 
        Our calculation shows
        a finite conductivity at zero temperature, which is consistent with
        experimental results. The region $kT < 0.02$ corresponds to the diffusive
        sector while for $kT > 0.05$ we are in the ballistic regime. 
	$W_c$ corresponds to the value of $W$ for which $\sigma_0$ at $T$ = 0 
	becomes non-zero and slowly starts to increase, while at $W = W_{m}$
	the system becomes a metal as the Hubbard gap is completely killed
	by the disorder. In between there is a narrow range of $W$ within 
	which $\sigma_0$ starts increasing from zero, but the system is still 
	however not a metal. 
Further increasing $W$, we enter a  metallic 
regime where the rate of fall in conductivity with temperature decreases.
This indicates that it is a highly disordered 
metallic phase (dirty metal).
The ground state of the highly 
disordered insulator is extremely difficult to obtain within Hartree Fock, 
which takes a very long time to converge to the required precision. Even when 
it does, there is a possibility of the 
system getting stuck in one of the many local minimas and not being able to 
access the true ground state. 
We therefore venture upto the point where there are no convergence problems(system is metallic). 

Though the effect of $U$ becomes stronger, with the Hubbard gap 
becoming larger as we decrease the dimensionality of the system, it is clear 
that disorder effect surpasses the effect of $U$ as one lowers the dimension as
evident from Fig.~\ref{fig.3}(d,e,f).
Fig.~\ref{fig.3}d,e,f shows the finite temperature plots of $\sigma_0$ for different 
parametric values of $W$ for $n_y$ = 40, 14 and 6 respectively. 
We again observe the general behaviour that the dc conductivity of an 
insulator/metal rises/falls with increasing temperature in the low 
temperature regime. However the onset of the metallic phase occurs for 
lower values of $W$ as we reduce the dimensionality.

We see that as we heat the system, 
the system goes from a low temperature metallic phase to a high temperature 
insulator phase. 
This effect is further seen to go away as we reduce the dimensionality.

\begin{figure}
  \begin{center}
    \begin{tabular}{cc}
      \resizebox{39.5mm}{!}{\includegraphics{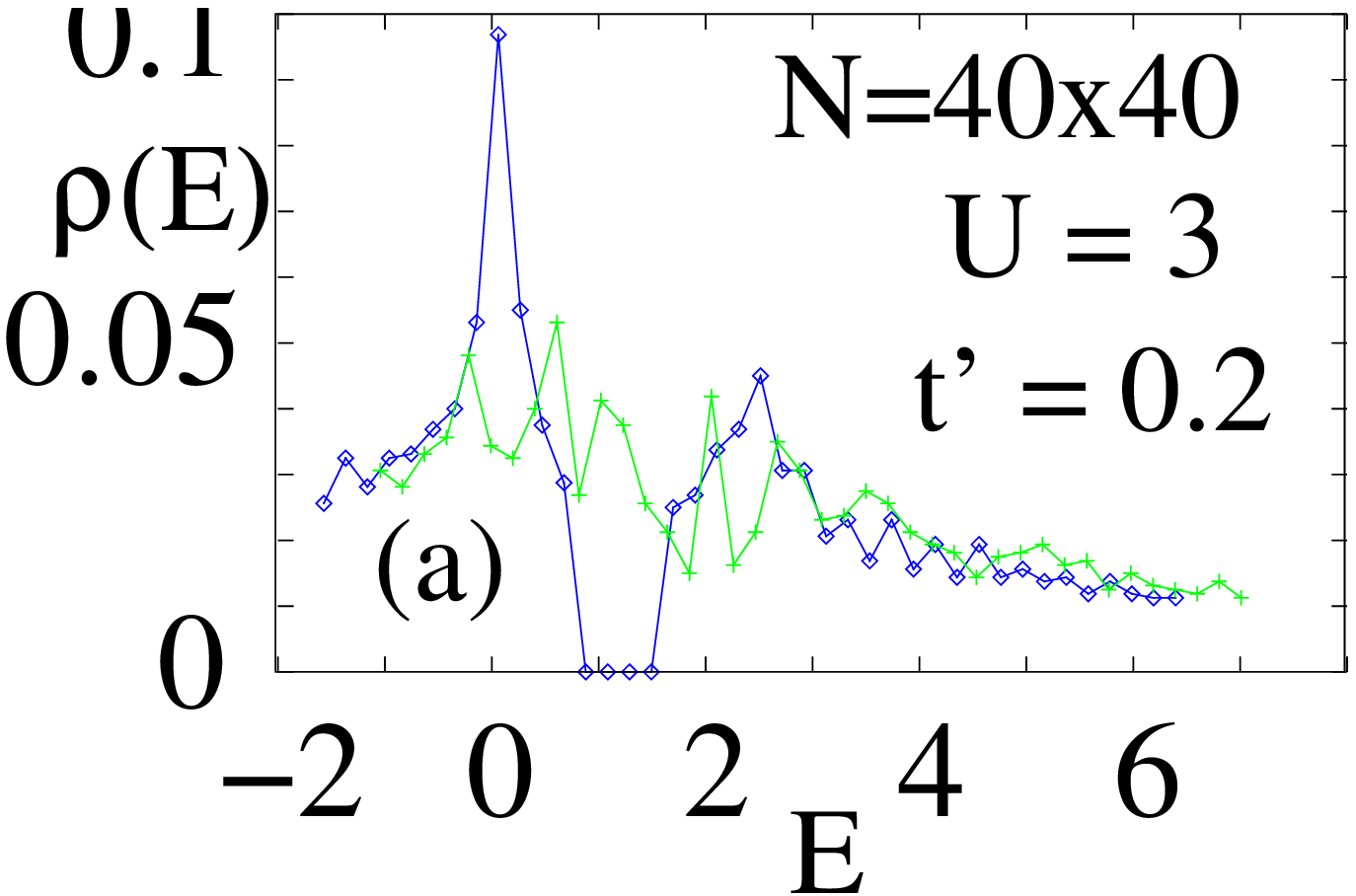}} & 
      \resizebox{39.5mm}{!}{\includegraphics{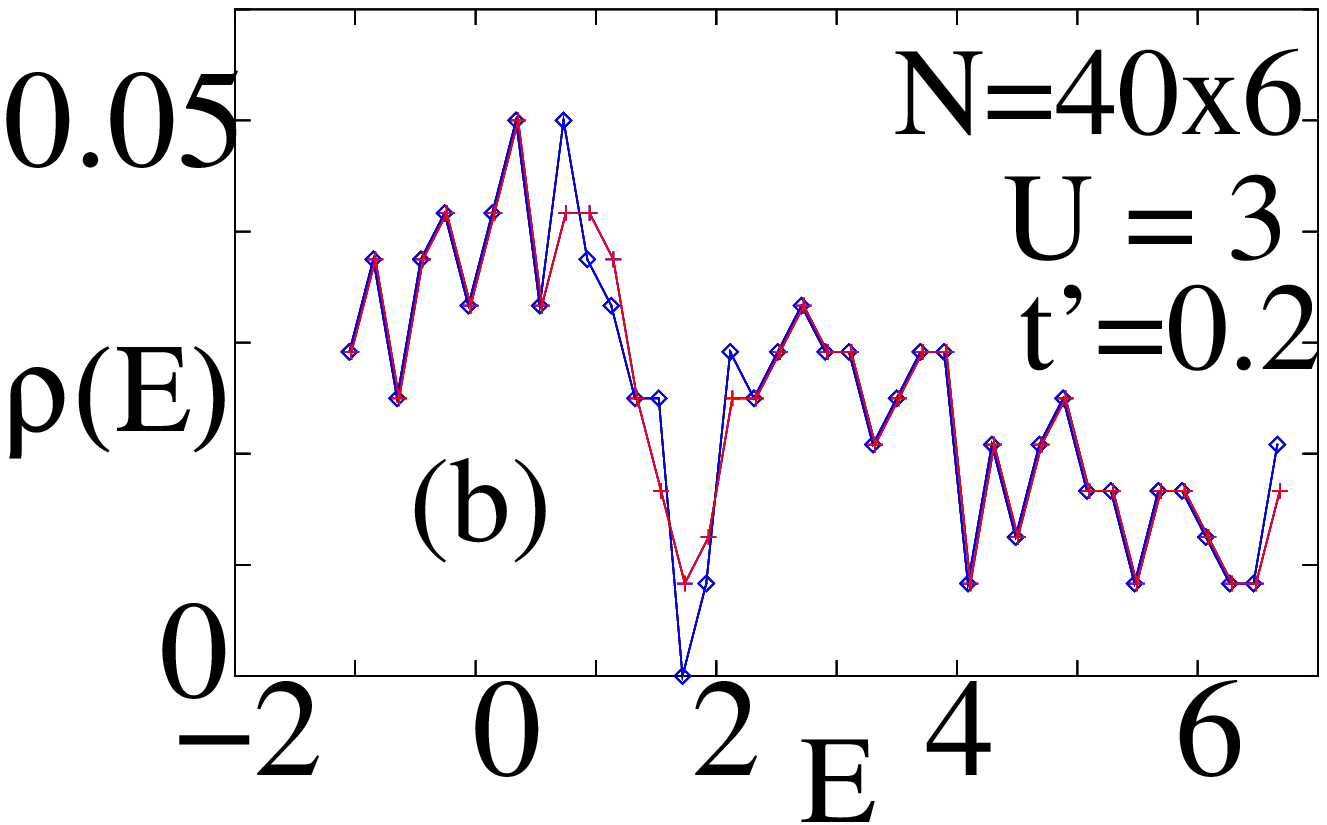}} \\
      \resizebox{39.5mm}{!}{\includegraphics{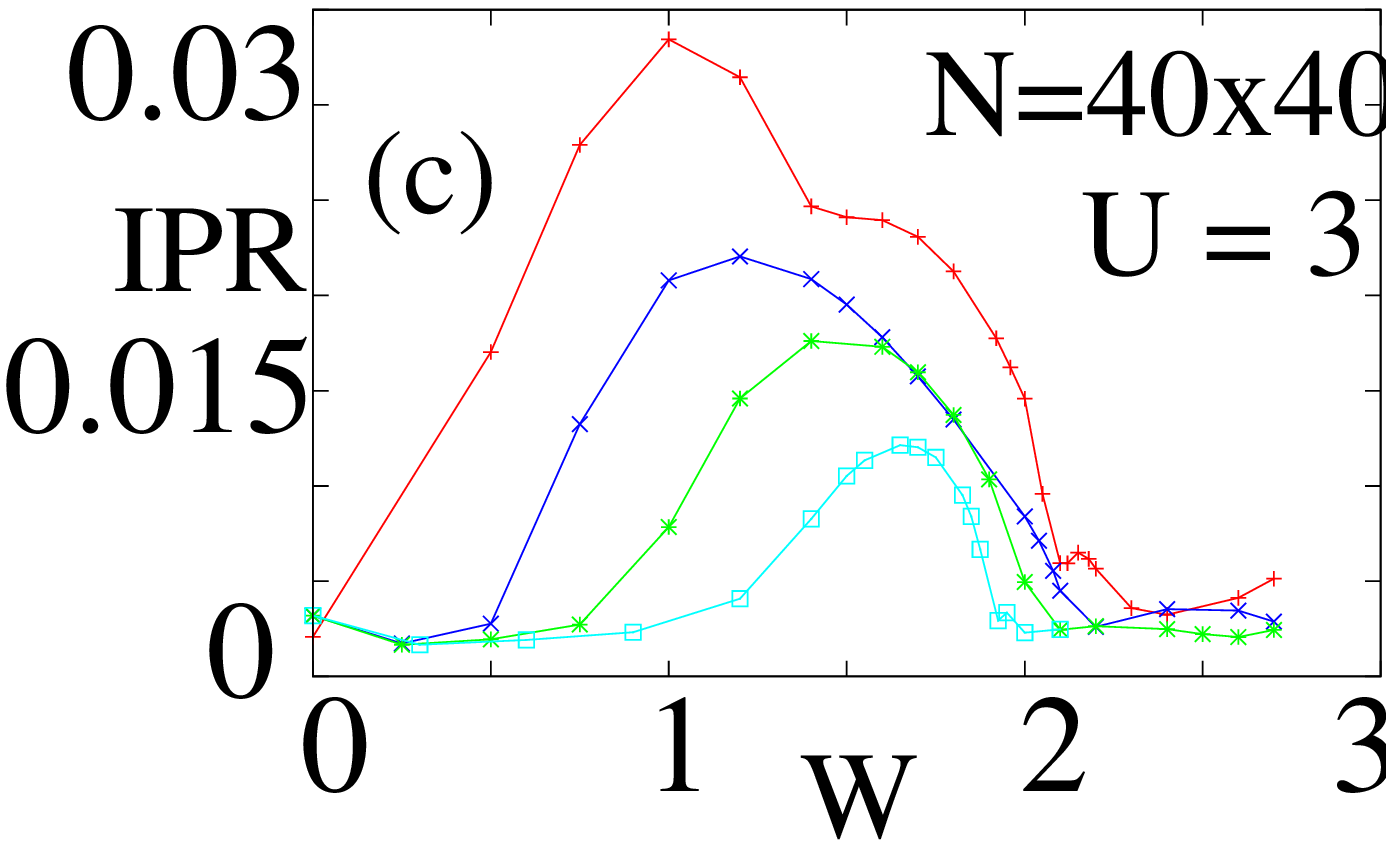}} &
      \resizebox{39.5mm}{!}{\includegraphics{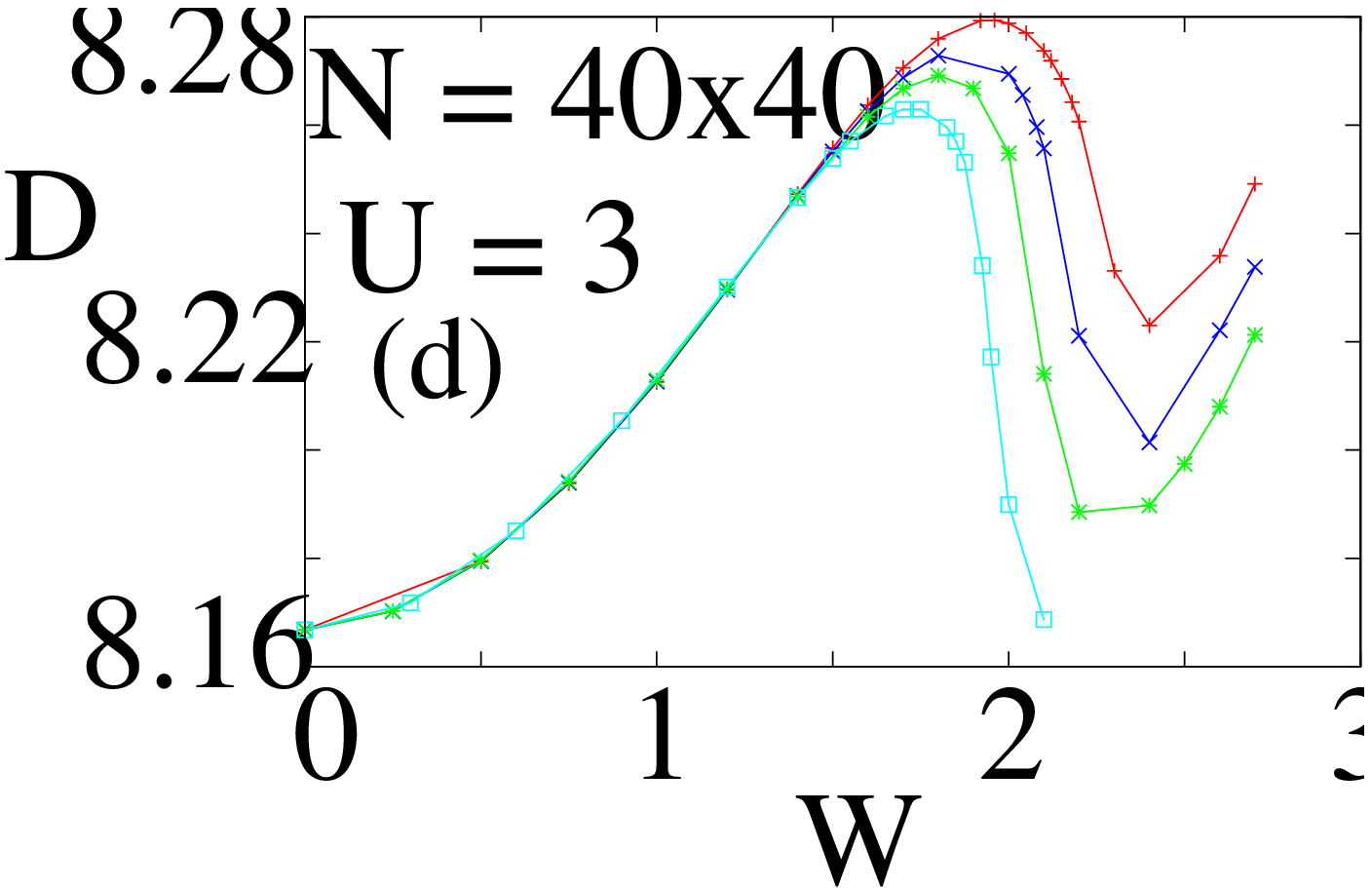}} \\
     \end{tabular}
\end{center}
\caption{ Plots of DOS in Fig.~\ref{fig.4}a and b,
IPR in Fig.~\ref{fig.4}c and BandWidth in Fig.~\ref{fig.4}d. The gap closes for $W$ = 
1.85 for $n_y$ = 40(Fig.~\ref{fig.4}a) whereas for $n_y$ = 6(Fig.~\ref{fig.4}b) it closes for $W$ = 1.6. 
The IPR and Bandwidth figs are for $t'$ = 0, 0.1, 0.15 and 0.2 respectively 
from top to bottom.}
\label{fig.4}
\end{figure}

Fig.~\ref{fig.4}a,b shows the plot of the density of states(DOS) for a fixed value 
of $U$ = 3 for $n_y$ = 40 and 6 respectively, highlighting how disorder closes 
the gap when we reach $W_m$. However the gap closes for lower value of $W$
as we decrease $n_y$ again emphasising the increased effectiveness 
of disorder as we reduce the dimensionality. 
We compare our result with the work done in the first reference of \cite{EffMediumDMFT}.
We observe from their LDOS plot that the MIT for the random disordered 2D Hubbard model 
occurs roughly at $W = 2.7$ for $U = 1.25$. In our work for the 2D $t-t'$ (for t'=0.2), 
deterministically disordered 
Hubbard model the MIT occurs at $W = 1.85$ for $U =3$ (Fig.~\ref{fig.4}a). 
Fig.~\ref{fig.4}c,d shows the IPR and the 
bandwidth of the system for a fixed value of 
$U > U_c$, as we increase the disorder strength. 
Fig.~\ref{fig.4}d clearly shows that for the interacting case the effect of 
disorder on band-width is quite non-monotonic and non-trivial. 
The bandwidth increases for very weak(low)
disorder, then it stops growing and surprisingly starts 
decreasing with increasing disorder. This is the regime where 
the states at the Fermi energy have already become 
delocalized as can be seen from Fig.~\ref{fig.4}c(the IPR curve also 
behaves non-monotonically). In this regime more  and more states are introduced 
in the gap region. Since total number of levels are fixed, the bandwidth decreases.  

The localization in the Hartree Fock single particle excitations for $t' = 0$, 
as reflected in the IPR peak below $W = W_m$ is rapidly eroded 
as we increase $t'$. This is seen in the rapidly 
falling peak structure in the IPR curve as we increase $t'$. 
Thus with increasing $t'$, there is a subtle transition 
in the nature of the wave-functions itself(localized to 
delocalized) as the metallic phase sets in. 

Fig 5(a) represents the plot of gap at half filling against disorder, for 
three different realizations of the Fibonacci sequence. One can see from 
these plots that the results remain almost invariant as we change the 
realization. The results shown here are for $U$ = 3 and $t'$ = 0.2 for 
a $40\times40$ system. 

Fig 5(b) shows our results of the dc conductivity at zero temperature 
as a function of disorder, in the range where the dc conductivity just 
starts to pick up from zero. We have shown the results for 2 system 
sizes, namely $32\times32$ and $40\times40$. It is clearly observed 
that the critical value of disorder $W_c$ where the dc conductivity 
picks up from zero is almost the same. This shows that the finite size 
effects are already quite small by the time we reach $32\times32$ system
size.  

\begin{figure}
  \begin{center}
    \begin{tabular}{cc}
      \resizebox{37.5mm}{!}{\includegraphics{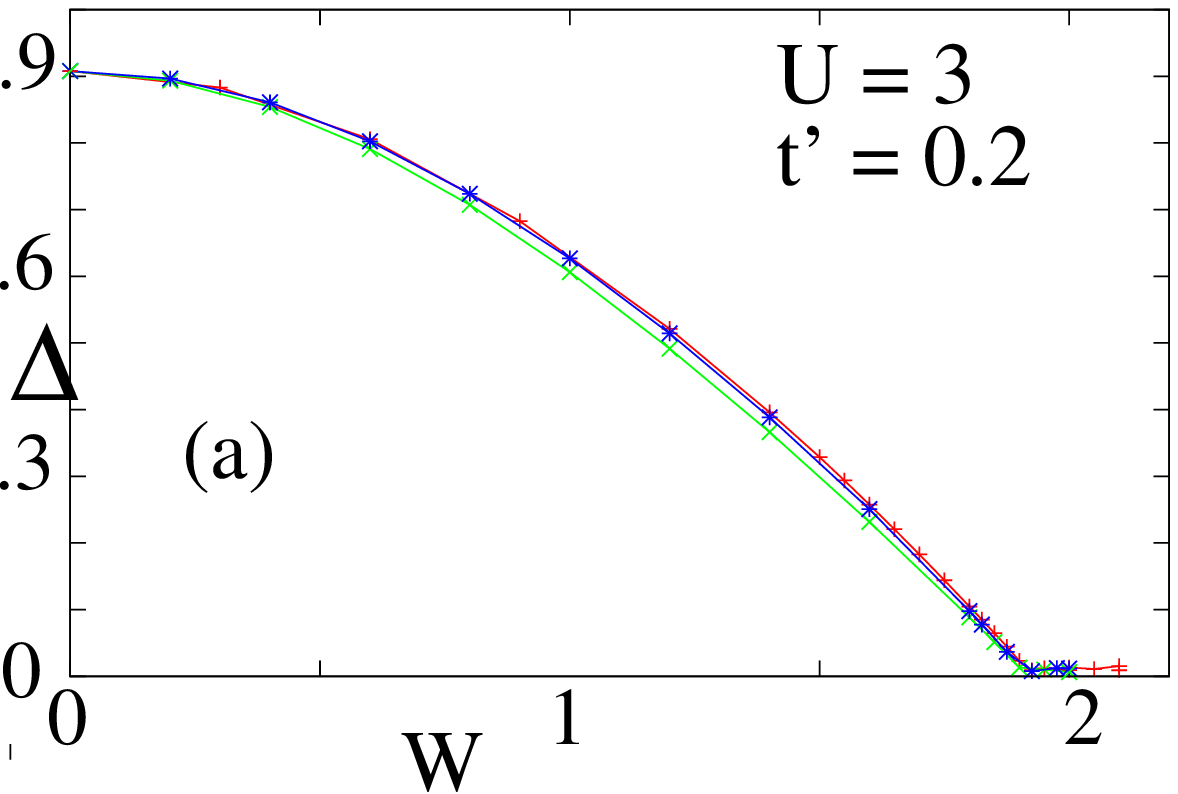}} & 
      \resizebox{37.5mm}{!}{\includegraphics{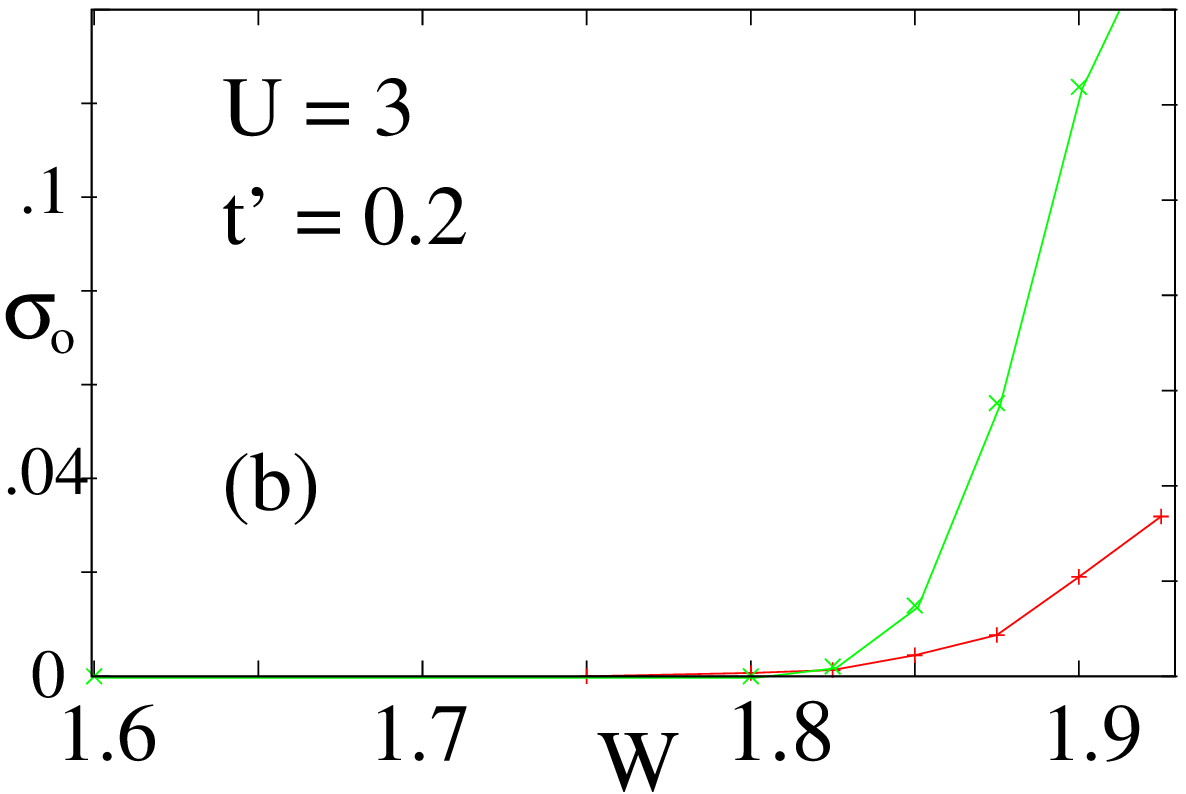}} \\
     \end{tabular}
\caption{Fig.~\ref{fig.5}a shows the effect of disorder on the gap at half
filling, for three different realizations of the Fibonacci sequence, for 
a $40\times40$ system. The uppermost, middle and lowermost curves correspond to 
the Fibonacci sequence beginning with first, seventh and third entries of the 
sequence respectively. 
Fig.~\ref{fig.5}b shows the plot of dc conductivity at $T = 0$ for $32\times32$(lower curve) 
and $40\times40$(upper curve) system size.  }
\label{fig.5}
\end{center}
\end{figure}

We believe that our simulation results provides useful and direct insight 
into the competition between correlations and disorder, which has been 
introduced in a simplified(deterministic) manner and is capable of capturing 
all the subtleties involved. It gives a good estimate of the 
Kubo conductivity as one goes from 2D to quasi 1D systems in presence 
of disorder and correlation and indicates the emergence of a metallic 
phase. As we tune the disorder to a very high value, the system becomes 
a metallic glass(highly disordered metal).  

\acknowledgments
Sanjay Gupta thanks Shreekantha Sil for some valuable suggestion. Tribikram thanks Saptarshi Mandal for providing additional 
computational facility while on a visit to IMSc and Sanjeev Kumar for 
useful discussions.

\end{document}